\documentclass[sigconf]{acmart}

\AtBeginDocument{%
  }

\acmYear{2022}\copyrightyear{2022}
\setcopyright{acmcopyright}
\acmConference[DroneCom '22]{ACM MobiCom Workshop on Drone Assisted Wireless Communications for 5G and Beyond}{October 17, 2022}{Sydney, NSW, Australia}
\acmBooktitle{ACM MobiCom Workshop on Drone Assisted Wireless Communications for 5G and Beyond (DroneCom '22), October 17, 2022, Sydney, NSW, Australia}
\acmPrice{15.00}
\acmDOI{10.1145/3555661.3560874}
\acmISBN{978-1-4503-9514-4/22/10}

\usepackage{listings}
\usepackage{amsmath}
\usepackage{algorithm}
\usepackage[noend]{algpseudocode}
\usepackage{caption}
\usepackage{subcaption}

\begin{document}

\title{BC-IoDT: Blockchain-based Framework for Authentication in Internet of Drone Things}

\author{Junaid Akram}
\affiliation{%
  \institution{The University of Sydney}
  \streetaddress{}
  \city{Sydney}
  \country{Australia}}
\email{jakr7229@sydney.edu.au}

\author{Awais Akram}
\affiliation{%
  \institution{COMSATS University Islamabad}
  \streetaddress{}
  \city{Vehari}
  \country{Pakistan}}
\email{awaisakram1212@gmail.com}

\author{Rutvij H. Jhaveri}
\affiliation{%
  \institution{Pandit Deendayal Energy University}
  \streetaddress{}
  \city{Gujarat }
  \country{India}}
\email{rutvij.jhaveri@sot.pdpu.ac.in}

\author{Mamoun Alazab}
\affiliation{%
  \institution{Charles Darwin University}
  \streetaddress{}
  \city{Darwin}
  \country{Australia}}
\email{Alazab.m@ieee.org}

\author{Haoran Chi}
\affiliation{%
  \institution{Universidade de Aveiro}
  \streetaddress{}
  \city{Aveiro}
  \country{Portugal}}
\email{haoran.chi@ua.pt}

\renewcommand{\shortauthors}{Akram et al.}

\begin{abstract}
  We leverage blockchain technology for drone node authentication in internet of drone things (IoDT). During the authentication procedure, the credentials of drone nodes are examined to remove malicious nodes  from the system. In IoDT, drones are responsible for gathering data and transmitting it to cluster heads (CHs) for further processing. The CH collects and organizes data. Due to computational load, their energy levels rapidly deplete. To overcome this problem, we present a low-energy adaptive clustering hierarchy (R2D) protocol based on distance, degree, and residual energy. R2D is used to replace CHs with normal nodes based on the biggest residual energy, the degree, and the shortest distance from BS. The cost of keeping a big volume of data on the blockchain is high. We employ the Interplanetary File System (IPFS), to address this issue. Moreover, IPFS protects user data using the industry-standard encryption technique AES-128. This standard compares well to other current encryption methods. Using a consensus mechanism based on proof of work requires a high amount of computing resources for transaction verification. The suggested approach leverages a consensus mechanism known as proof of authority (PoA) to address this problem . The results of the simulations indicate that the suggested system model functions effectively and efficiently. A formal security analysis is conducted to assess the smart contract's resistance to attacks.
\end{abstract}

\ccsdesc[500]{Security and privacy~Authentication}

\keywords{IoDT, Access Control, security, consensus, Cluster Head}

\maketitle

\section{Introduction}
The Internet of drone things (IoDT) refers to the infrastructure currently being developed to allow anyone to control and access drones over the internet \cite{hussain2021amassing}. In reality, drones are becoming more common consumer commodities, enabling any user to do a range of tasks in regulated airspace. Despite the fact that technology has enabled mass manufacture of onboard UAV components like CPUs, sensors, storage, and battery life, the performance constraints of these components have decreased expectations\cite{akram2021swarm, akram2018energy}. With IoDT, drones and cloud mobility capabilities may be paired, allowing for remote access and operation, scalability in offloading, and cloud storage of data acquired from a great distance.

Because IoDT can gather environmental data, several companies, such as those dealing with energy trading, surveillance, smart grids, etc, may profit from it \cite{akram2021cloud,akram2020intelligent, wang2021sparse}. It connects to the Internet and automates the monitoring infrastructure without assistance from a third party. The drones of the IoDT network are responsible for environmental monitoring. However, drones in IoDT encounter obstacles such as non-repudiation, restricted resources, and the existence of malicious nodes.  Numerous solutions \cite{kar2021cl, akram2022using} are presented as potential solutions to the aforementioned problems. However, difficulties like single point of failure (SPOF) and performance bottlenecks develop owing to the centralized structure of these studies.

Using blockchain technology, researchers have developed a variety of solutions to remove the need for intermediaries in IoDT, hence eliminating the aforementioned difficulties. Blockchain technology is a secure, distributed ledger that avoids, among other possible problems, centralized authority and third-party participation.  Additionally, blockchain's usage of a distributed and immutable ledger helps alleviate trust difficulties between organizations that were previously unknown. As a critical component of the network architecture, miners validate the legitimacy of all transactions conducted by nodes. Miners use several consensus methods, including  proof of work (PoW), proof of authority (PoA), proof of stake, to validate the authenticity of these transactions. Proof of work (PoW) and proof of authority (PoA) are two examples. In the PoW method, each network node contributes to the solution of the mathematical problem. The first node to answer the challenge is responsible for transaction verification. First, the transactions are validated for correctness before being added to the blockchain. Moreover, blockchain makes use of smart contracts, the purpose of which is to finalize all rules and conditions. In addition, it removes the need for an intermediary. In addition, blockchain ensures network security by evaluating the trust between nodes, and by detecting malicious nodes using the Merkle tree \cite{goyat2020blockchain}.

Multiple blockchain-based solutions for scalability concerns, high costs, and single points of failure have been suggested \cite{cui2020hybrid}. However, blockchain is a highly expensive storage method, and all of these networks store their data there. Currently, it costs \$14,151.68 to store one megabyte of data on a blockchain network. In addition, a proof-of-work consensus mechanism is used in \cite{cui2020hybrid}, which is inappropriate in a low-resource context.

In the context of IoDT networks, routing refers to the process by which individual nodes establish connections with one another and transfer data from its source to its eventual destination. Different nodes within the IoDT, including drones, cluster heads (CH), and base stations (BSs), supervise the transfer of data . It is the responsibility of the CHs to do any required data processing before transmitting it to the BSs \cite{haseeb2019intrusion, mehmood2018efficient, khan2021spice, xu2021mobile}. However, there are no authenticated network nodes. Since any node may get network access, any harmful action is conceivable. Moreover, data storage on a blockchain is costly since no cost-effective storage strategy is presented in \cite{haseeb2019intrusion}. Since blockchains are intended to record transactions in perpetuity, the problem of inadequate capacity to store these data quickly becomes apparent. In an IoDT configuration, an excessive energy drain on CHs may also affect network performance. In \cite{cui2020hybrid}, no new CH selection processes are suggested. In order to add new blocks to the blockchain and validate transactions, PoW needs miners to solve this problem. This mining procedure is time-consuming due to the intricacy of the underlying problem, which increases the network's computational expenses over time. 

The main contributions of this work are as follows:

\begin{itemize}
    \item IPFS is used for IoDT's distributed storage requirements. A payment scheme is recommended as an incentive for archival use of IPFS. 
    \item To exclude untrustworthy nodes, each node's identification is verified. The suggested LEACH technique, named "R2D", identifies and selects CHs from drones using the least distance, highest degree, and maximum residual energy.
    \item Blockchain technology is used for secure service provisioning systems, with the suggestion that an advanced symmetric encryption technique with a 128-bit key be employed to protect sensitive data.
    \item For the purpose of determining the validity of the smart contract, a formal security analysis is undertaken.

\end{itemize}

The paper moves on to the subsequent sections. In Section 2, the proposed system model is provided. Section 3 presents the simulation outcomes and system model validation. In Section 4, the formal security analysis is addressed. The sixth section provides an overview of the accomplished activities.

\begin{figure*}
	\centering
	\includegraphics[width=5in]{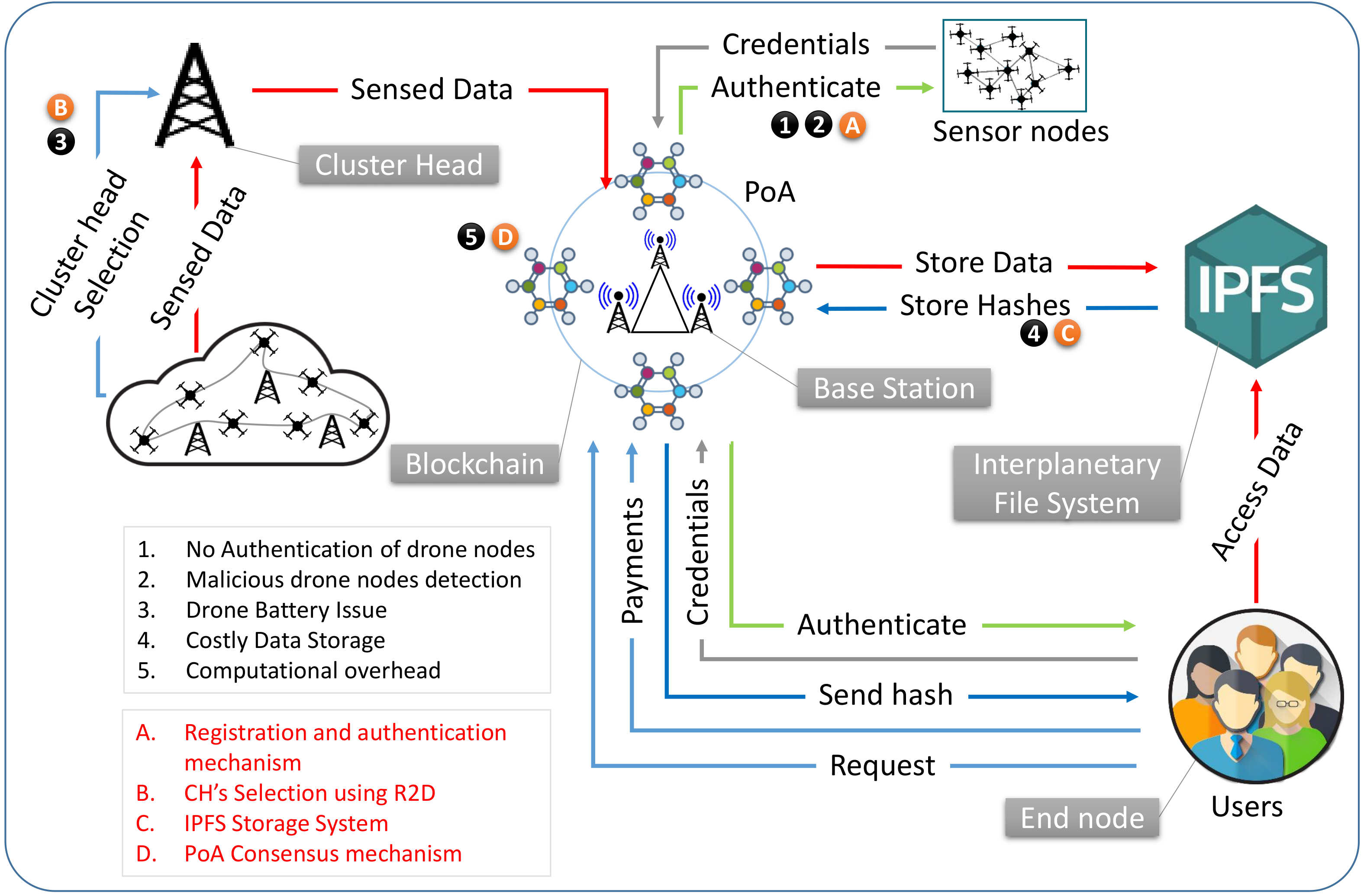}
	\caption{Drone nodes’ Authentication and Cluster Head Selection in IoDT based on Blockchain.}
	\label{AuthBlock}
\end{figure*}

\section{System Model}

Figure 1 depicts the system model, which will be analyzed and reviewed below. IoDT, end-users, the Interplanetary File System (IPFS), and a blockchain are system components. The ultimate shift of drones' backends to IoDT will be enabled by technologies such as the Internet of Things, intelligent computer vision, cloud computing, enhanced wireless connectivity, big data, and cutting-edge security. A growing variety of industries are embracing the use of drones, including agriculture and industry, government and commercial organizations, and the monitoring of smart cities and rural areas. Each user must register for the blockchain network and undergo the authentication process to guarantee the network's security. This objective is motivated by the usage of registration and authentication procedures \cite{cui2020hybrid}. IPFS is a distributed system that utilizes data and information chunks for storage and retrieval. Any time information is saved on IPFS, a hash is generated. The 32-bit hash is a result of the IPFS technique for storing data. The blockchain is then updated with this hash.

\subsection{Initialization}
A smart contract is a computerized contract whose conditions may be carried out with little or no requirement for a third party. Blockchain technology makes this capability possible. It is deployed on the transaction-controlling BSs of the network. Utilizing blockchain technology, sensor node credentials are kept in a blockchain for subsequent authentication, as part of the process of enrolling nodes in the network. The credentials are sent using a message like the equation shown below.

\begin{equation}
   Message = Packet(ID_N, MAC_N, Rep_N)
\end{equation}

Drone Node ID ($ID_N$), MAC addresses ($MAC_N$), and reputations ($Rep_N$) are utilized as part of the registration and authentication procedure for nodes. $Rep_N$ is a measure of a node's network standing that takes into account the node's past behaviors and its connections to other nodes and the rest of the system. If the node's reporting is not honest, its reputation will decrease, but if it is, it will grow. The credentials are recorded in an asymmetrical manner on a blockchain-based distributed ledger. Their addresses are also recorded on the blockchain, which provides an additional degree of security against fraudulent system usage. Throughout the authentication procedure, the credentials are compared to the information already on blockchain. When credentials do not match previously saved data, the node in question is marked as  malicious.

\subsection{Cluster Head Selection}
Drone nodes, cluster heads, and base stations make up the IoDT's infrastructure. The drone nodes transmit data to the CHs, where computations are done. CHs must first gather drone node data before relaying it to base stations (BSs). When BS wants to store data, it makes a request to IPFS, which responds with a hash of the data. During data processing and storage, the available energy of CHs is quickly depleted. Our suggested solution to this problem is based on a technique known as R2D, which chooses CHs from regular nodes. CHs is determined by three factors: available spare energy, proximity to BSs, and maximum node degree. If there is a single node that meets all of the given requirements, it will be classified as the CH. It is responsible for combining all data packets before transmitting them. If many nodes are found to match the requirements, CH will be chosen at random.

\begin{algorithm}[t]
\caption{Cluster Head Selection}
\begin{algorithmic}[1]

    \State \textbf{Inputs:} Deployment of N nodes, BSs and CHs
    \State \textbf{Outputs:} Cluster Head
    \State Select maximum energy node from set N
    \State Select the node having minimum distance from BS
    \State Select maximum degree node from set N
    \For{ i=1 to N}
    \State Select $(Max (S(i).Er), Max (S(i).Degree) and Min (S(i).Distance))$
    \If{New selected $CH = Max(Degree, Er)$ \&\& $Min(Distance)$}
    \State Selected node is CH
    \EndIf
    \State Check next node
    \EndFor
    
\end{algorithmic}
\end{algorithm}

\subsection{Authentication}

Drone nodes can only be trusted if they have been verified by the blockchain. By identifying and eliminating malicious sensors, authentication guarantees that only trustworthy nodes are used to provide services. The security of your registration information is improved when you send it via an encrypted connection. At the node level, the data is encrypted using the public key of the Blockchain Storage node. The Blockchain Storage node will use its private key to decode the information before adding it to the blockchain. There is no new way for transferring encryption and decryption keys since they cannot be canceled or altered. This ensures a steady functioning of the system. When a node in the network does any kind of activity, the BS checks to make sure it has the proper authorization to do so. Network-wide security may be guaranteed using this method. Multiple research papers \cite{cao2021iibe,abdi2021laptas} depend on node authentication. The use of encryption for node authentication is outlined in detail in \cite{cao2021iibe}. Centralized authority results in bottlenecks and decreased productivity. To verify users, \cite{abdi2021laptas} employs a simple token system. The origin of SPOF may be traced back to authoritative monopolies. The sink node is acknowledged by the nodes as a means of authentication. The sink node is identified by the nodes based on their position in the sequence. The use of Proof-of-Work (PoW) for node authentication raises computational expenses. Hybrid blockchain nodes are validated in \cite{cui2020hybrid}. The use of PoA raises the price of calculation. Keeping data on a blockchain network might be rather expensive. For the sake of lowering the overall computational cost, the system model we gave makes use of PoA consensus for node authentication. The presented authentication model has been seen before. This authentication framework provides a safe place to store sensor data and a robust method for selecting a suitable CH. Since authentication processes are used in several contexts, this approach represents a new combination \cite{cui2020hybrid, abdi2021laptas}. A lower total network cost is achieved via the use of IPFS's distributed data storage. Authentication takes place when the credentials used by different nodes to access the blockchain are found to be a match.

To perpetrate crimes, an attacker node may move information across the network once it has been taken over. All valid nodes have verified their identity with BSs, and their registration information is recorded in the distributed ledger. Since authentication information is not recorded in the blockchain, any malicious node that is part of the data transfer may be quickly identified and kicked from the network. This prevents malicious nodes from taking over any other node in our network and doing criminal acts. After being disconnected from the network, a rogue node can no longer send or receive information. Because the blockchain is loaded on them, BSs are able to confirm transactions and cut off the rogue node's access. The suggested architecture employs Proof of Authority (PoA) for transaction mining. It requires less processing power than proof-of-work consensus methods. PoA miners are chosen nodes in the network in advance. As soon as an authentication request is made, registered nodes are checked. The blockchain record consist of three attributes: $ID_N$, $MAC_N$, and $Rep_N$. In order to determine whether or not two nodes are compatible, the blockchain checks the credentials of each node against the credentials that have already been entered in the blockchain. If the credentials are correct, the nodes may be trusted and their legitimacy can be announced. They may be misunderstood as malicious or insincere. Algorithm 2 describes the procedure in detail.

\subsection{Data Storage}
As a result of the high cost of keeping data on a blockchain, this huge volume of data cannot be kept there. Data is sent from BSs and stored in IPFS with hash values and the credentials of blockchain nodes. All financial dealings are recorded chronologically on a blockchain ledger. This ledger is shared across all entities in the network. Each node plays a vital role in maintaining the network's integrity. After a transaction occurs, all ledgers are immediately synced. When a malicious node tries to alter a ledger's transaction record, it's easy to spot since the altered ledger won't be consistent with any other ledgers. IPFS saves information in blobs. Data stored on IPFS is not kept permanently. If data is to be kept on IPFS for an extended period of time, cost has to be paid. Data storage on the IPFS is incentivized by monetary rewards for participating peer nodes. The blockchain keeps a record of hashes. For security reasons, the given hash may be used by only approved nodes. IPFS is a temporary storage solution. We provide IPFS an incentive to keep data around for a while.

To authenticate and register nodes, we use a hosted private blockchain. Requests to register as a user are recorded on the blockchain. Then, blockchain verifies the nodes' registrations. When a user is registered, the system ignores their request. If not, then node registration is permitted. The authentication procedure is described in detail in Algorithm 2.

\begin{algorithm}[t]
\caption{Drone Authentication}
\begin{algorithmic}[1]

    \State \textbf{Inputs:} $ID_N, MAC_N, Rep_N$
    \State \textbf{Outputs:} Nodes authentication message
    \State Registration← Input ($ID_N, MAC_N, Rep_N$)
    \If{$ID_N, MAC_N, Rep_N$ stored in blockchain}

    \State Node authenticated
    \EndIf
    \If{$ID_N, MAC_N, Rep_N$ are not stored in blockchain}
    \State Node not authenticated
    \EndIf
    \State Recommend for registration
\end{algorithmic}
\end{algorithm}

\subsection{Service Provisioning}

Blockchain can verify the legitimacy of a user's identification documents. Transactions cannot take place until a user's identity has been verified on the blockchain. For each node, the blockchain verifies the credentials to make sure they are not duplicates. If the provided credentials are consistent with those on file, then the user is considered to be legitimate. If they don't match, the node is considered malicious and is removed from the network. After verifying the user's identity, we next make sure the amount of ethers being purchased is sufficient. The request will be denied and the hash of the data will not be sent if the user does not have enough ethers. users will have access to network services upon authentication. When asked, BS will encrypt the service using the customer's private key and give the ciphertext.

Network security is handled via 128 bit AES encryption. Data integrity is protected by both AES encryption and SHA-256 hashing. The BS will generate the hash using SHA-256 and add it to the blockchain. The BS encrypts the information using the client's private key and transmits the result. Information is encrypted and sent, then the user uses the sender's private key to decrypt the information. The user decrypts the data, then computes the hash using SHA-256 and checks it against the hash stored on the blockchain. Together, SHA-256 and AES ensure the integrity and privacy of your data. Our primary focus is on providing timely information to users, and AES 128-bit encryption allows for efficient encryption and decryption.

\section{Results and Discussion}
The simulated networks have three kinds of nodes: 100 drone nodes, 4 cluster heads (CHs), and 2 base stations (BSs), all of which are distributed throughout a 1000x1000$m^2$ area. In addition, our system's model includes the user's address with the sensed data and the data index. Drone nodes dispersed across the network are responsible for data collection. The information obtained by these sensors is sent to cluster heads for processing. After initial processing, the data is sent to the BSs for further analysis. The BS stores both the data and the index to that data. BSs store the data index and the discovered information as part of their service. In addition, the user's address is the Ethereum address that is assigned to the transaction at the time it enters the network. After registering, nodes are screened to verify they are not malicious and can contribute to the network.

\begin{figure}
     \centering
     \begin{subfigure}[b]{0.48\columnwidth}
         \centering
         \includegraphics[width=\textwidth]{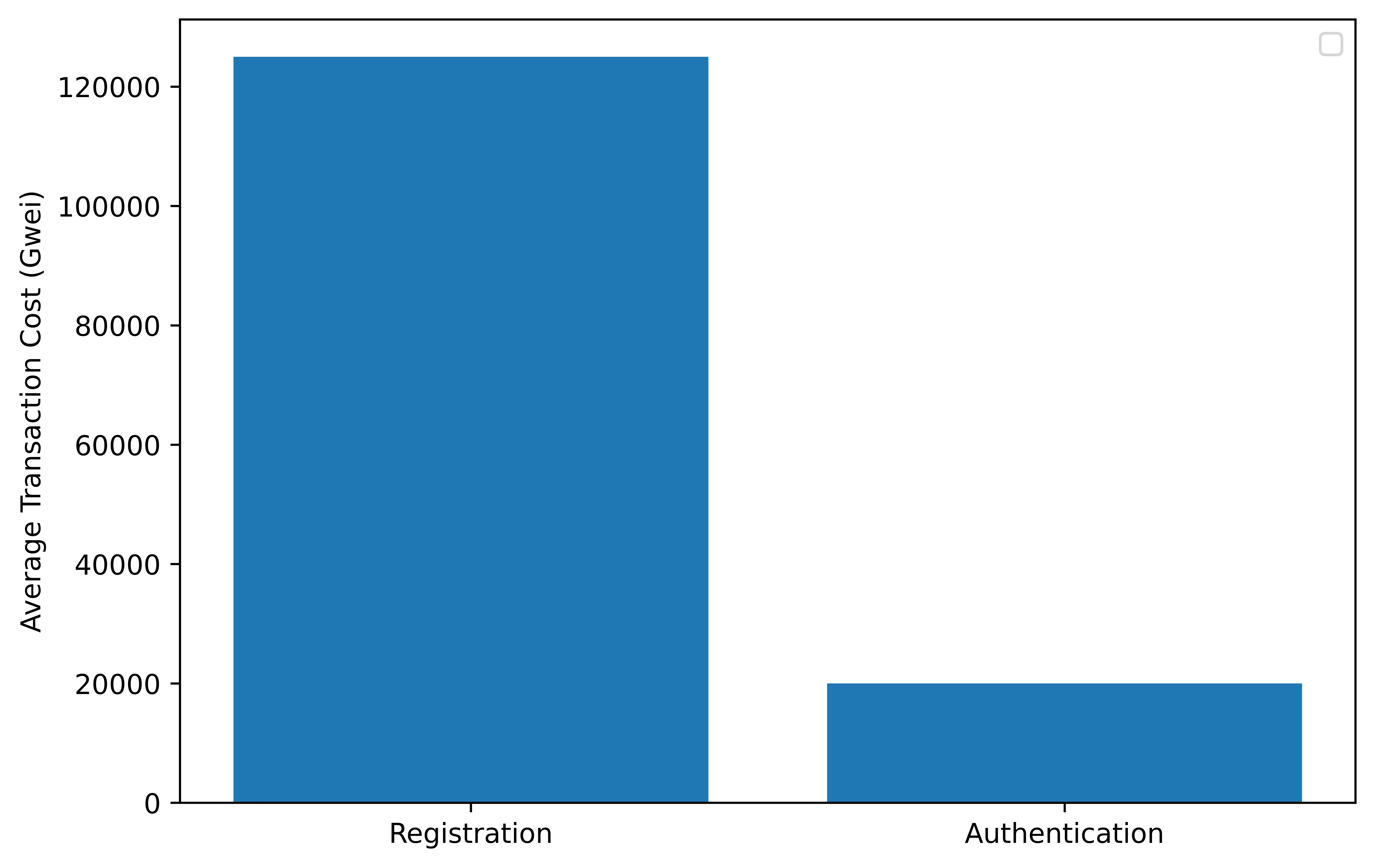}
         \caption{}
         \label{regauth}
     \end{subfigure}
     \hfill
     \begin{subfigure}[b]{0.48\columnwidth}
         \centering
         \includegraphics[width=\textwidth]{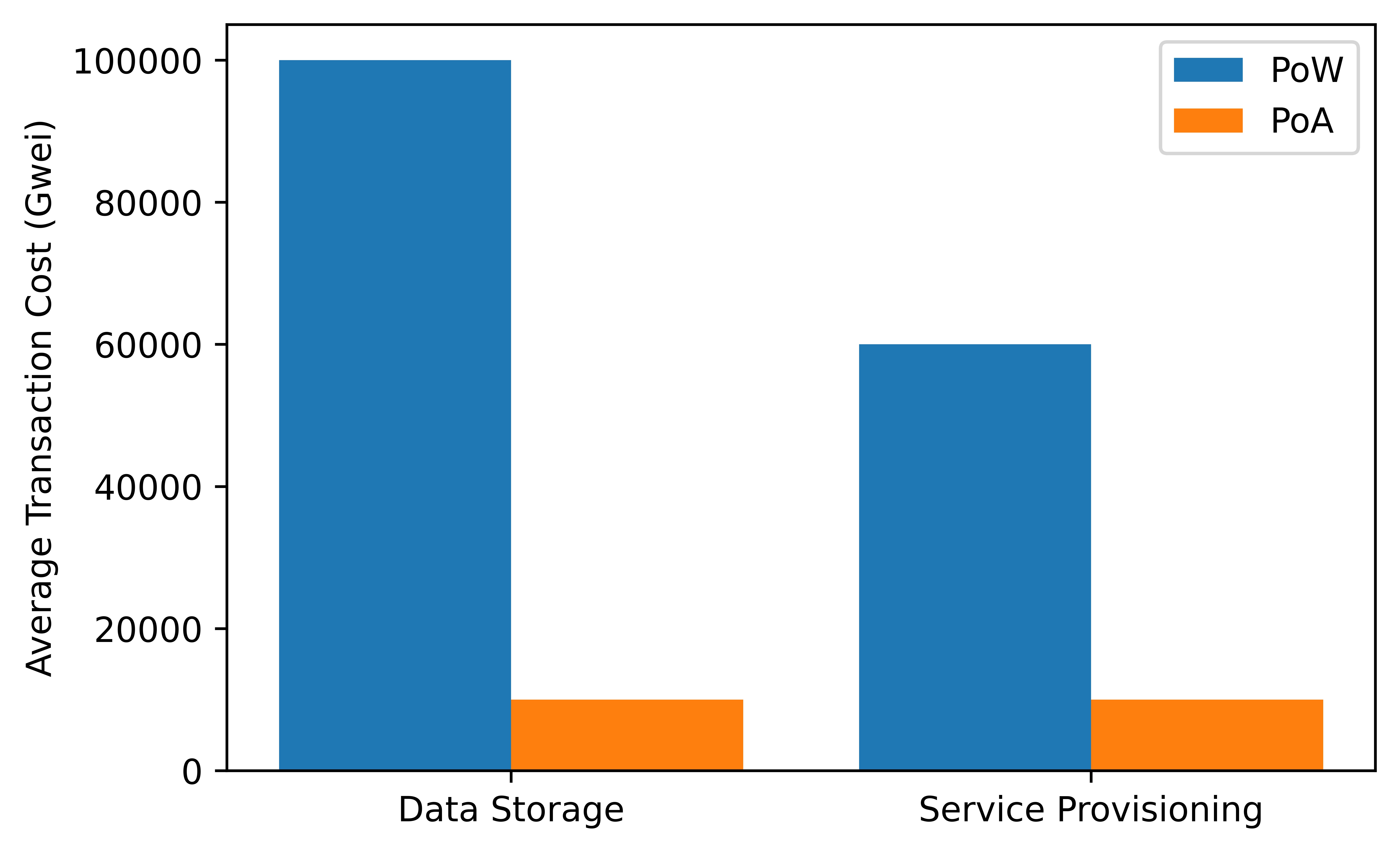}
         \caption{}
         \label{trancost}
     \end{subfigure}

        \caption{(a) Node registration and authentication Transaction cost. (b) IPFS data storage and service provisioning transaction costs.}
        \label{fig:three graphs}
\end{figure}

Figure \ref{regauth} depicts the total transaction costs incurred throughout the registration and authentication procedures. A transaction fee is paid for each use of a smart contract. The findings indicate that the transaction cost increases as the number of packets grows. The transaction charge will grow according to the network's size. Registration information must be validated during the authentication process in order to save transaction costs. Figure \ref{trancost} displays a comparison between the Proof-of-Work and Proof-of-Authority consensus procedures as they pertain to the delivery of services and the storage of data. The proposed system uses a private blockchain, which makes it possible to perform the PoA consensus procedure. This helps to the overall cost reduction of the proposed system. The Gwei-based computing expenses of the two consensus techniques are evaluated. Proof of Authority is superior to Proof of Work in terms of performance. According to PoA, mining nodes are not needed to do any difficult mathematical computations. In PoA, miners are pre-selected, and it is their duty to check the validity of all transactions. This explains why PoW incurs far higher computational costs than PoA. Figure \ref{trancost} depicts the creation of a smart contract based on the PoA consensus technique when users want blockchain services. This happens when a client launches a blockchain deployment. Comparing the two service delivery methods reveals that PoA uses less computing resources. Figure \ref{trancost} demonstrates how the average IPFS transaction cost is determined when data is stored in a distributed file system. As previously noted, the average transaction cost of PoW is higher than that of PoA.

\begin{figure}
     \centering
     \begin{subfigure}[b]{0.48\columnwidth}
         \centering
         \includegraphics[width=\textwidth]{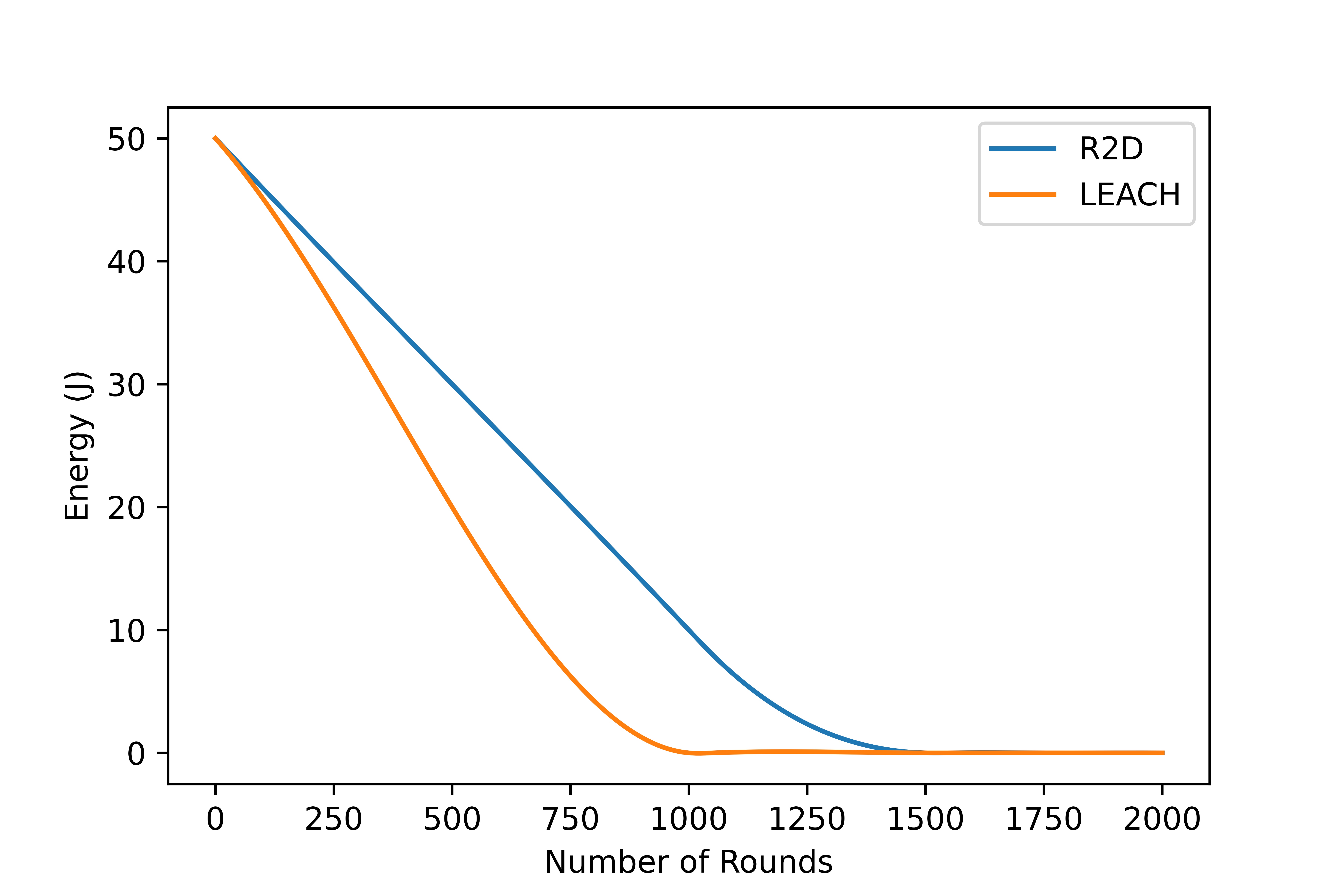}
         \caption{}
         \label{enc}
     \end{subfigure}
     \hfill
     \begin{subfigure}[b]{0.48\columnwidth}
         \centering
         \includegraphics[width=\textwidth]{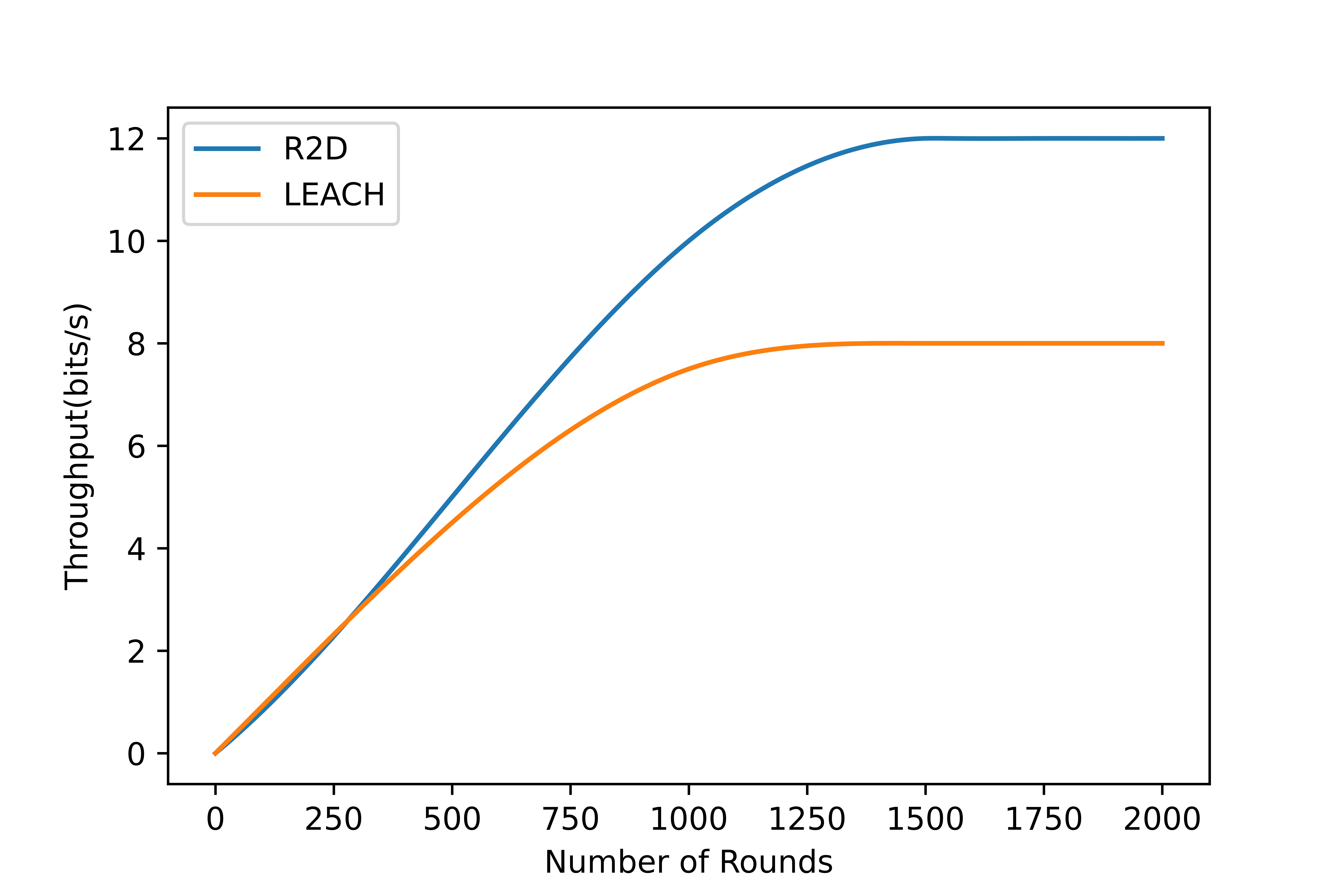}
         \caption{}
         \label{thr}
     \end{subfigure}

        \caption{(a) Energy consumption. (b) Network throughput.}
        \label{fig:three graphs}
\end{figure}

Figure \ref{enc} compares the total number of rounds to the total amount of energy used. We examine the energy efficiency of the LEACH protocol and R2D. The nodes of R2D will continue to use energy throughout the length of 1400 cycles. In LEACH, the nodes continue to consume energy until the completion of the 1000th cycle. LEACH's overall energy usage is high due to its random selection of CHs. There is a possibility that the sensor nodes closest to the base station (BS) will not join the network or be designated as cluster head. When selecting which CHs to employ, the R2D considers three criteria: maximum degree, minimum distance, and minimum energy usage. Figure \ref{thr} demonstrates that the initial network throughput for both protocols is zero. The throughput of the network increases with rounds. Since each node contributes to R2D, the quantity of data delivered from ordinary nodes to base stations continues to increase. It starts to stagnate at 1500 rounds. When R2D is used to choose CHs, throughput increases and the number of available nodes for data packet transmission is maximized.

\section{Security Analysis}
A formal examination of network security may identify potentially dangerous nodes. Using this authentication technique, malicious nodes may be identified and eventually eliminated. The network's toughness is measured using attacks like Sybil and Man in the Middle (MITM). Oyente is a tool that examines the identification data of network nodes as recorded in the blockchain. Below are evaluations of attacks.

\textit{Sybil attack:} Removal of malicious nodes is carried out by the authentication process. A network node's registration and authentication is done by the authentication system which is a prerequisite for the drone to perform any operation. Drone data is recorded and stored in a blockchain. The blockchain keeps track of the verified identities of each node in the network and their respective identifiers. This attack is impossible in the proposed framework because of the two-way nature of the authentication.

\textit{MITM attack:} Network conversion is vulnerable when MITM attack is induced to R2D . Because of how the nodes handle authentication, attacks like these are impossible. Credentials saved in a blockchain are used by nodes to join the network. These nodes need to be validated before they may begin interacting. The authenticity of the node is verified by comparing it against registration data stolen by the adversary. The registration process generates a random key for each node. This key is used for authentication purposes. The given key can be found on the blockchain. A malicious node trying to join the network would have to submit the same credentials as a legitimate one. During the registration process, the attacker node is removed. The attacker node needs the unique identity stored on blockchain in order to authenticate. In order to ensure the accuracy of its information, the attacker must use the key.

Man-in-the-middle attacks occur when an adversary in the communication chain either delivers false data or makes unauthorized changes to a legitimate data packet. When malicious actors intervene, information routed from drones to CHs or from CHs to BSs is lost. A constant stream of malicious packets is sent to the victim by the attackers. Each drone node is assigned a different key ID. The connected nodes in a network must first verify the identity of each other via authentication. The attacker node in a Sybil attack pretends to be other nodes in the network. During an attack, only malicious packets are transmitted to the target, which significantly degrades the network's performance. If rogue users can be identified by mutual registration and authentication, network performance can be enhanced.

Attacks on a network use a higher amount of energy. Sending malicious packets to the target site consumes more resources. Following registration and validation, the network comprises only trusted nodes. The IoDT network can save energy when there are no threats to deal with. As a means of attacking other nodes in a network, attackers in the network may often send out corrupted data packets. They also consume a lot of power. This prevents valid nodes' data packets from reaching their intended destinations. Authentication will result in the removal of malicious nodes. Since genuine nodes consume less energy, they live longer. In the attacker model, malicious nodes quickly die because they run out of energy while sending out the corrupted data packets. 

\begin{table}
	\centering
	\caption {Smart contract vulnerability analysis using oyente tools}
	\small
	\begin{tabular}{  c  c  }
		\hline
		\textbf{Vulnerability} & \textbf{Result}\\  \hline
		
		EVM Code Coverage & 99.5\% \\
		Integer Underflow & False \\
		Integer Overflow & False \\
		Parity Multisig Part 2 & False \\
		Callstack Depth Attack & False \\
		Transaction-Ordering Dependence & False \\
		Time Stamp Dependency & False \\
		Re-entrancy  & False \\
		
\hline

	\end{tabular}
	\label{oyente}
\end{table}

\subsection{Smart Contract Analysis}
Table \ref{oyente} depicts the Oyente tool's study of the node registration and authentication smart contracts' security. The table demonstrates that the smart contract can survive the attacks outlined in the article. These results disprove every one of the presented attacks against the smart contract.

Strategies that helped us to tackle smart contract vulnerabilities are as follows.

\begin{itemize}
    \item \textit{Re-entrancy attack:} A malicious network node will invoke the external function whenever the smart contract is executed. Due to this, the true function will not operate as effectively. Our smart contract is resistant to this kind of attack since all network nodes must first be authorized. The blockchain records the verifiable identities of each network node. Consequently, invalid nodes are gradually eliminated from the network since they cannot offer relevant data.

    \item \textit{No double spending:} In our prototype system, clients verify their identities using a private key that we provide. No malicious node can access the network's information, hence the network is safe.

    \item \textit{Denial of service attack:} Our system model is resistant to a denial-of-service attack because the users are authenticated by delivering the communication's secret keys and because they are required to exchange the keys before obtaining data.

    \item \textit{Single point of failure:} IPFS's decentralized nature is essential to the architecture of our system. The inflexibility of the centralized data store prevents it from responding rapidly. Our suggested remedy makes this attack obsolete with IPFS, a decentralized file-sharing system. The technology is quick and responsive while storing and retrieving data.
\end{itemize}

\section{Conclusion}
We propose a blockchain and IoDT-based network aimed to reduce the malicious activities while reducing the required resources. Authentication mechanisms are used to guarantee that only authorized nodes can communicate. Only nodes with network access credentials are permitted to submit data for processing to CHs. CHs collect and analyze data. We propose the R2D protocol in which CHs are chosen among regular drone nodes based on their highest degree, smallest distance from BS, and largest residual energy. CHs with low energy are replaced with nodes that meet the above requirements. Additionally, the collected data are saved in the IPFS database. Blockchain technology is used for long-term storage. Services are sent via a secure connection utilizing a 128-bit AES-encrypted protocol. We may infer, based on the simulation findings, that data storage and service supply can be achieved with minimal computational cost. Moreover, the transaction costs associated with registration and verification are small. Comparing LEACH to R2D, researchers discovered that LEACH delivers higher energy efficiency, longer network life, and higher throughput. The results demonstrate that R2D greatly outperforms LEACH. A formal security analysis is conducted to assess the smart contract's resistance to assaults. To demonstrate the network's security, it is also subjected to the Man-in-the-Middle attack (MITM) and the Sybil attack. Future initiatives will use a machine learning method to identify malicious drone nodes.

\bibliographystyle{ACM-Reference-Format}
\bibliography{sample-bibliography.bib}

\end{document}